\documentclass[]{aa}
\usepackage{epsf,psfig,epsfig,graphics}
\begin{document}
    \title{Low frequency radio monitoring of Cygnus\,X-1 \\and Cygnus\,X-3}

    \author{M. Pandey\inst{1}, A. P. Rao\inst{2}, G. G. Pooley\inst{3}, P. Durouchoux\inst{4},
      R. K. Manchanda\inst{5}, C. H. Ishwara-Chandra\inst{2}}

    \offprints{M. Pandey : mamta@ncra.tifr.res.in}

    \institute{Dept. of Physics, Mumbai University, Mumbai - 400 098, India\and
NCRA, TIFR, Post Bag 3, Ganeshkhind, Pune - 411 007, India\and
Cavendish Laboratory, University of Cambridge, Madingley Road, Cambridge CB3 0HE, UK\and
CNRS FRE 2591/CEA Saclay, DSM/DAPNIA/SAP, F-91191 Gif sur Yvette Cedex, France\and
Tata Institute of Fundamental Research, Mumbai-400005, India\and
}

    \date{Received ; accepted}
    \authorrunning{M. Pandey et al.}
    \titlerunning{Low frequency radio monitoring of Cygnus\,X-1 and Cygnus\,X-3}

\abstract{
We present results of monitoring observations of the micro-quasars
Cygnus\,X-1 and Cygnus\,X-3 at 0.61 and 1.28\,GHz.
The observations were performed with Giant Meter-wave Radio Telescope,
GMRT, between 2003 June and 2005 January.
Variable, unresolved sources were found in both cases.
Cyg\,X-1 was detected in about half of the observations,
with a median flux density about 7\,mJy at each frequency. 
The results show clearly that there is a break in the
spectrum above 1.28\,GHz.
The variations in Cyg\,X-1 may be due to refractive interstellar scintillation. 
Cyg\,X-3 was detected in each observation, and varied by a factor of 4.
For this source, models of the scintillation suggest a very long timescale
(of the order of 40 yr at 1.28\,GHz), and therefore we believe that the
variations are intrinsic to the source.

}

\maketitle

\section{Introduction}
The X-ray sources Cygnus\,X-1 and Cygnus\,X-3 are accreting X-ray binary systems
which have been classified as micro-quasars. The X-ray emission in such systems 
is believed to originate in the inverse Comptonization of seed photons in the 
accretion disc; the sources exhibit several spectral states classified as low-hard, 
intermediate and high-soft state  (Cui 1999). The spectral analysis of the
radio emission from these X-ray binaries with  black hole candidates (BHCs) suggests 
that the radio emission arises due to synchrotron emission from the high energy 
electrons emitted in the jets (Hjellming \& Johnston 1988; Hjellming \& Han 1995)
and the radio emission is quenched when a source is in high-soft X-ray state 
(Corbel et al. 2000), and detected in both low-hard and intermediate X-ray state.
High-resolution mapping of X-ray binaries, particularly the well-known object GRS\,1915+105, 
has demonstrated the presence of relativistic outflows (hence the term `micro-quasars'). 
Such motions have been mapped in Cyg\,X-3 (Mioduszewski et al 2001), and a jet-like 
feature has been mapped in Cyg\,X-1 (Stirling et al 2001). There has been a variety of 
other studies, mostly at high radio frequencies (above 2\,GHz). However, the low frequency 
characteristics of Cyg\,X-1 in particular have not been explored so far.\\ To understand 
the physical mechanism connecting the inflow and outflow of matter
in such systems, a comprehensive study of the correlation between radio and X-ray
emission in different states of these sources is necessary.
We have carried out systematic  monitoring of micro-quasars using the Giant
Meter-wave Radio Telescope, GMRT (Pandey et al. 2004) at low frequencies.
In this paper, we present the observations of Cyg\,X-1 and Cyg\,X-3, which demonstrate 
variations in the flux densities at low radio frequencies for both these sources.
The  observed flux density variations can be ascribed either to intrinsic
mechanisms specific to each source, i.e. a process occurring around the compact sources 
themselves, or to refractive interstellar scintillation. The refractive scintillation 
can be described as  the flux density variations caused by ``focusing'' and ``defocusing'' 
of electromagnetic waves by large scale plasma density irregularities in the 
interstellar medium  (Rickett et al. 1984).

\section{Observations and Analysis}
The radio snapshot observations (of duration 30 min) were carried out at 0.61\,GHz and 
1.28\,GHz with a bandwidth of 16/32 MHz using the Giant Meter-wave Radio Telescope, GMRT.
The flux density scale was set by observing the primary calibrators 3C286 and 3C48.
Phase calibrators were interleaved with 25 min scans on each of Cyg\,X-1 and Cyg\,X-3.
The sampling time was 16 s. The data recorded from GMRT was converted into FITS files 
and analyzed using the Astronomical Image Processing System (AIPS). A self calibration on 
the data was used to correct for phase related errors and improve the image quality.
Tables 1 and 2 summarize details of observations using GMRT at the two frequencies between
June 2003 and Jan 2005. Cyg\,X-3 was detected in each observation and Cyg\,X-1 during about 
one-half of them. Columns 1, 2 and 3 in both the tables gives the dates of observations and 
the measured flux densities or upper limits. Column 4 gives the  corresponding rms noise in 
the field of the radio image. For the observations performed during telescope maintenance/test 
time the background noise was higher because of the reduced number of antennas available.
For each field there are some 6 background sources detected. We record in Tables 1 and 2 the 
measured flux density of one from each field, at an angular distance of 15$'$ (Cyg\,X-1) and 
22$'$ (Cyg\,X-3). The measured flux densities of these `control' sources are consistent
with constant actual flux densities and the observed noise levels. This result gives us 
overall confidence in the reliability of the system.\\
To supplement the data from GMRT, we have also used radio data on Cyg\,X-1 and Cyg\,X-3 taken 
with the Ryle Telescope at 15\,GHz. The longitudes of the GMRT and Ryle Telescope differ by 
$74\deg $ (5 hr), and so the observations are usually not exactly simultaneous; in the worst 
cases the nearest 15-GHz observation is separated by 2 days from the GMRT observation. 
The radio light curve at 15\,GHz for Cyg\,X-1 was averaged for 10 min, that for Cyg\,X-3 for 
5 min. The typical uncertainty is 2\,mJy + 3\% in the flux scaling. In column 6 of the Tables 
we have listed  the flux density at 15\,GHz for the nearest observation at that frequency.
Column 7 gives the spectral indices between the GMRT observation and 15\,GHz (in the sense 
$S \propto \nu^{\alpha}$). Column 8 gives  the X-ray spectral state of the source as determined 
from the 2-12 keV light curves of the sources  taken from the ASM archival data.

In Fig. 1 and Fig. 2 we show the first ever radio images of Cyg\,X-1 and Cyg\,X-3 at 0.61\,GHz.
The AIPS subroutine \texttt{jmfit} shows that the images are consistent with point sources.

\begin{table*}[t]
\centering
\caption{Flux density of Cyg\,X-1 between 2003 Jun and 2005 Jan}
\begin{tabular}{llllp{7mm}p{8mm}lll}
\hline
\hline
\multicolumn{1}{c}{Date} &\multicolumn{2}{c}{$S_\nu$ }&GMRT rms&Control& Source&$S_\nu$  &$\alpha$&X-ray\\
\multicolumn{1}{c}{MJD}  &\multicolumn{2}{c}{(mJy)}&noise (mJy)&\multicolumn{2}{c}{(mJy)} &(mJy) &&state\\
$\nu$/GHz &0.6 &  1.2   &                              &0.6  & 1.2    &15 &        &  \\
\hline
52796&  -     & $\le$4.8&   1.6   &-     &4.2      &13     &$\ge$0.6   &H-S\\
52799&  -     &  10.1   &   1.3   &-     &5.2      &22     & 0.3       &H-S\\
52827&$\le$2  &      -  &   0.5   &9.1   &-        &8      &$\ge$0.4   &H-S\\
52829&$\le$1.5&      -  &   0.4   &8.9   &-        &13     &$\ge$0.5   &H-S\\
52845&  -     & $\le$7  &   2.2   &-     &4.7      &6      &$\ge$0.06  &H-S\\
52848&  -     & $\le$5  &   1.3   &-     &4.8      &6      &$\ge$0.07  &H-S\\
52855&  -     &   5.4   &   0.2   &-     &5.7      &22     & 0.8       &Int\\
52891&  5.8   &      -  &   0.7   &9.2   &-        &15     & 0.3       &L-H\\
53037&  -     &   9.2   &   0.2   &-     &4.4      &15     & 0.2       &Int\\
53102&  9.7   &      -  &   0.4   &10.3  &-        &22     & 0.3       &L-H\\
53104&$\le$7  &      -  &   2.0   &8.9   &-        &27     &$\ge$0.4   &L-H\\
53107&  10.3  &      -  &   0.7   &10.3  &-        &26     & 0.3       &L-H\\
53127&$\le$7  &      -  &   2.1   &10.5  &-        &16     &$\ge$0.3   &L-H\\
53377&$\le$5.1&      -  &   1.7   &8.7   &-        & 6     &$\ge$0.05  &H-S\\
53379&$\le$3.3&      -  &   1.1   &8.5   &-        & 6     &$\ge$0.2   &H-S\\
53391&  8.4   &      -  &   0.9   &9.5   &-        &20     &     0.3   &Int\\
53392&  8.4   &      -  &   0.8   &9.3   &-        &16     &     0.2   &Int\\
\hline
\end{tabular}
\end{table*}

\begin{figure}[h]
\begin{center}
\psfig{figure=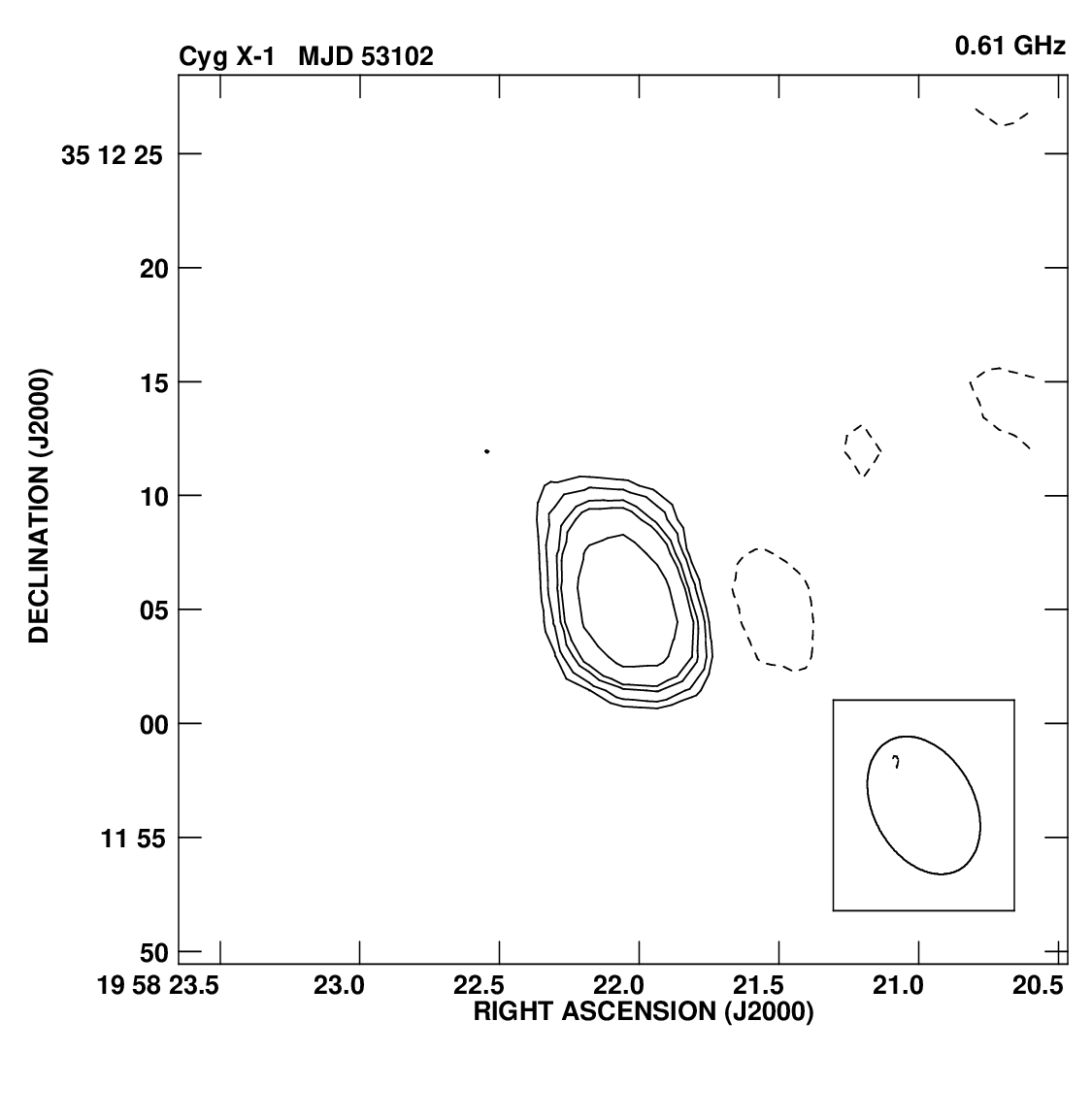,angle=0,width=8.5cm}
\end{center}
\caption{ GMRT image of Cyg\,X-1. The contour levels are
$1.3 {\rm mJy} \times (-1, 1, 1.4, 2, 4, 8)$.}
\end{figure}

\begin{figure}
\begin{center}
\psfig{figure=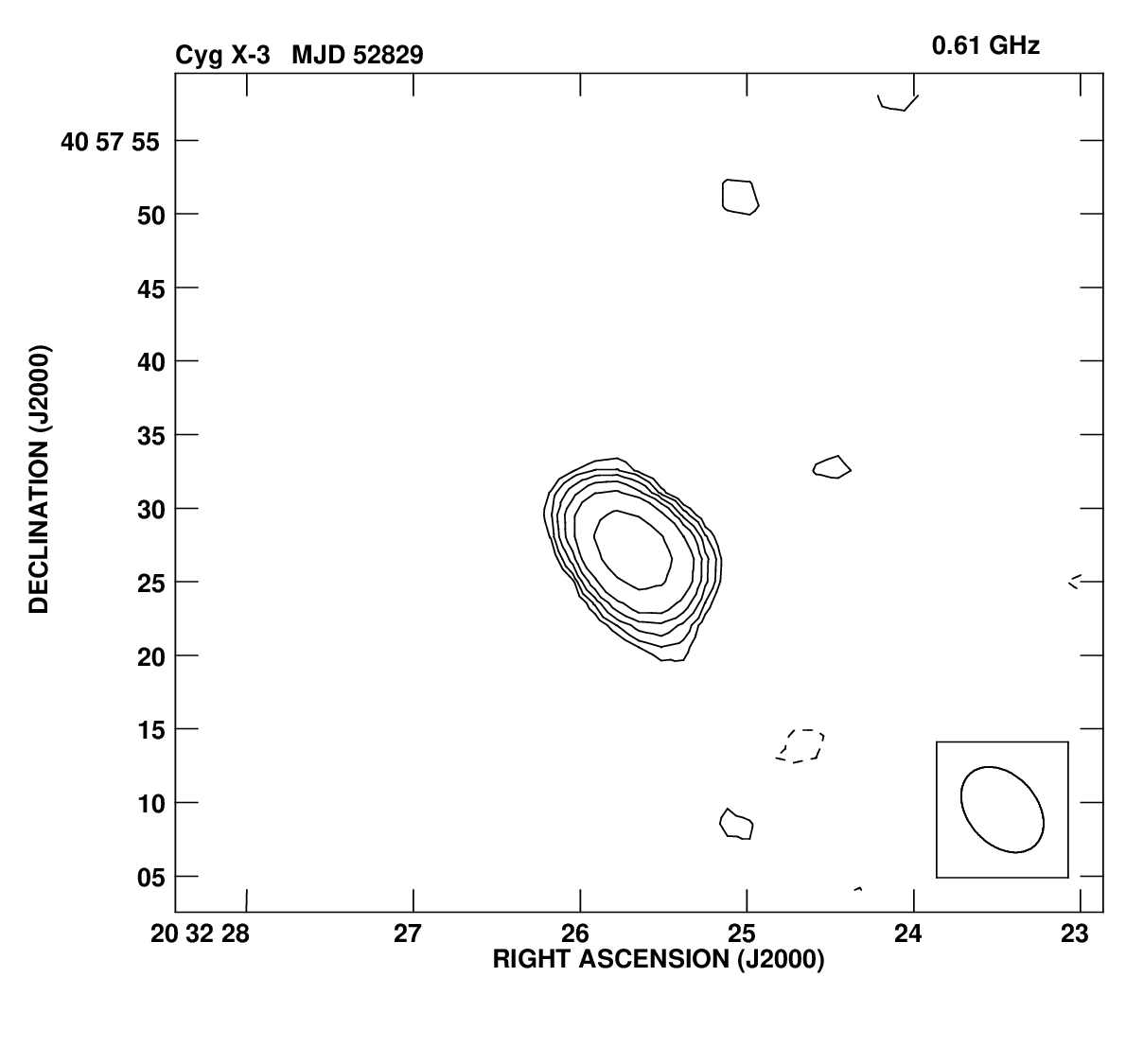,angle=-90,width=8.5cm}
\end{center}
\caption{{ GMRT image of Cyg\,X-3.
The contour levels are $2.3 {\rm mJy} \times (-1, 1, 1.4, 2, 4, 8)$}}
\end{figure}

\begin{table*}[t]
\caption{Flux density of Cyg\,X-3 between 2003 Jun and 2005 Jan}
\centering
\begin{tabular}{llllp{7mm}p{8mm}lll}
\hline
\hline
\multicolumn{1}{c}{Date}&\multicolumn{2}{c}{$S_\nu$} & GMRT rms&Control& Source&$S_\nu$ &$\alpha$&X-ray\\
\multicolumn{1}{c}{MJD}&\multicolumn{2}{c}{(mJy)}&noise (mJy)&\multicolumn{2}{c}{(mJy)}&(mJy)& &state\\
$\nu$/GHz &0.6 &  1.2   &                              &0.6  & 1.2    &15 &        &  \\
\hline
52797&  -    &  57.9  &   0.4   &-      &7.7       &70           &0.08     &L-H\\
52800&  -    &  90.4  &   0.5   &-      &9.6       &110          &0.08     &L-H\\
52830&31.2   &      - &   0.8   &16.6   &-         &115          &0.4      &L-H\\
52849&  -    &  62.4  &   0.4   &-      &9.4       &190          &0.5      &L-H\\
52856&  -    &  53.5  &   0.3   &-      &8.4       &132          &0.4      &L-H\\
52892&7.0    &      - &   0.7   &17.8   &-         &73           &0.7      &L-H\\
53037&  -    &  21.2  &   0.3   &-      &8.1       &51           &0.4      &L-H\\
53105&14.5   &      - &   1.1   &16.6   &-         &45           &0.4      &L-H\\
53107&13.0   &      - &   0.8   &20.1   &-         &70           &0.5      &L-H\\
53127&40.8   &      - &   1.1   &18.6   &-         &50           &0.06     &L-H\\
53128&29.5   &      - &   0.8   &19.5   &-         &73           &0.3      &L-H\\
53378&31.4   &      - &   1.7   &18.7   &-         &130          &0.4      &H-S\\
53379&29.8   &      - &   1.1   &18.5   &-         &130          &0.5      &H-S\\
53392&27.3   &      - &   0.9   &19.5   &-         &170          &0.6      &H-S\\
53393&44.2   &      - &   0.8   &19.3   &-         &170          &0.4      &H-S\\
\hline
\end{tabular}
\end{table*}

\section{Discussion}
\subsection{Cyg\,X-1}
Cyg\,X-1 is  bright X-ray source with a mass in excess of
7$M_{\sun}$ (Gies \& Bolton 1986) and a distance of $\sim$2.5 kpc.
The companion is an O9.7\,Iab super-giant (V1357 Cyg, HDE 226868),
showing a strong stellar wind (Persi et al. 1980) and almost filling
its Roche lobe (Bolton 1972).
The observed radial velocity for the system $v \sin i$  is about
76 km s$^{-1}$ (Gies \& Bolton 1982).
\begin{figure}[ht]
\begin{center}
\psfig{figure=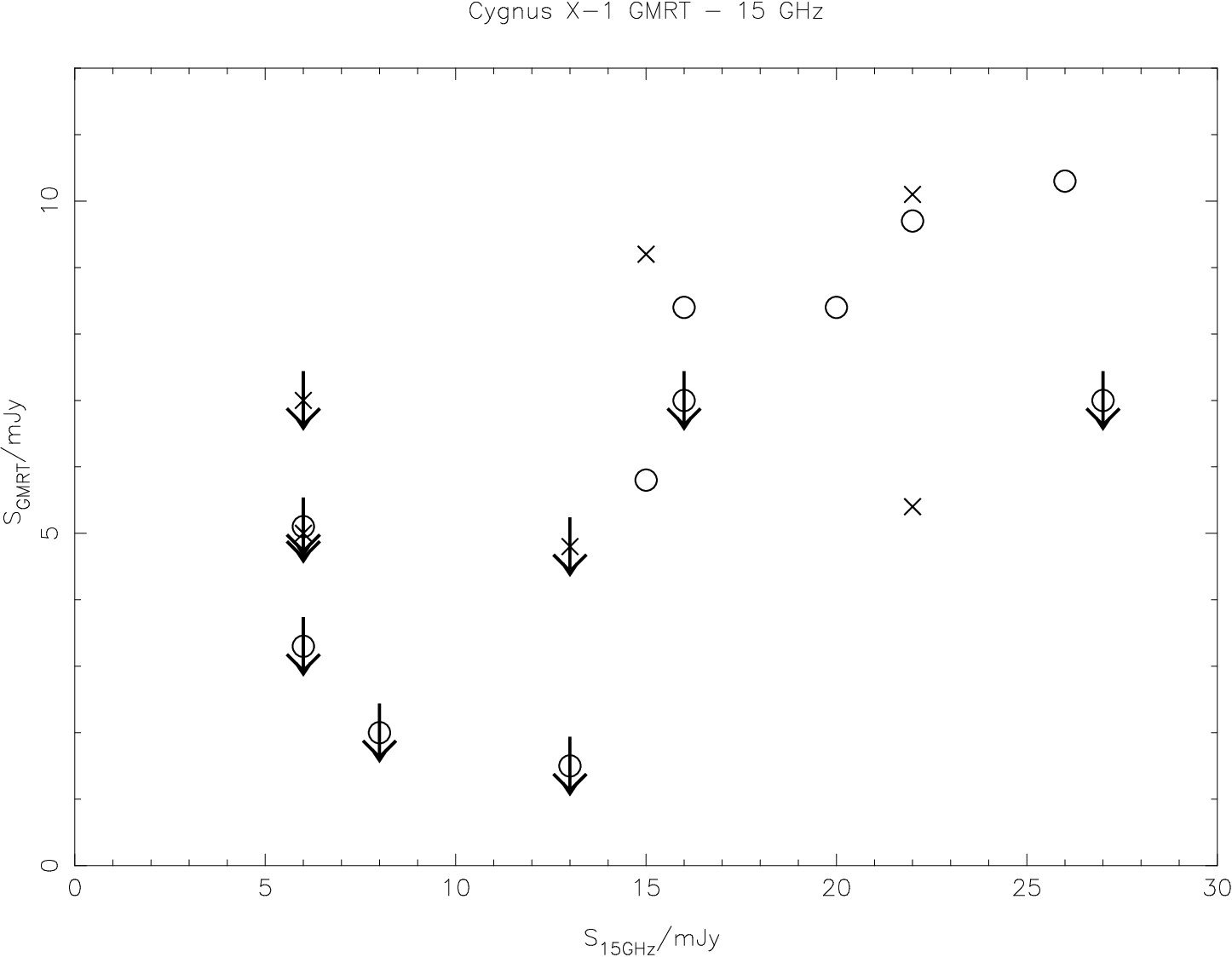,angle=-90,width=9cm}
\end{center}
\caption{Flux--flux plot for Cyg\,X-1 showing the low-frequency
GMRT flux densities ($\bigcirc$ 0.61\,GHz, $\times$ 1.28 GHz)
and the closest 15-GHz measurement.}
\end{figure}
The emission in X-rays and in the radio band vary over a wide range of
time-scales, including modulation at the 5\fd 6 orbital period;
see, for example, Brocksopp et al (1999) and Gleissner et al (2003).
The X-ray emission is usually in the spectral state
described as `low-hard' (with a non-thermal
component extending to a few hundred keV); the relatively rare
`high-soft' state is dominated by a strong low-energy component.
The high-frequency radio emission tends to be suppressed during the
high-soft states (Hjellming, Gibson \& Owen 1975) (see Fig. 4).
While the X-ray source is in the low hard state, the radio emission in
the cm-wave band is persistent, but the magnitude does vary  by a factor of about 5,
with very rare outbursts to higher flux densities. These variations are much
less spectacular than those of some other members of this class such as Cyg\,X-3
and GRS 1915+105. The radio spectral index $\alpha$, defined in the sense
$S \propto \nu^{\alpha}$, is close to zero in the range 2 -- 220 GHz
(Fender et al 2000). Jet emission in Cyg\,X-1 has also been observed at milliarcsec resolution
during the low-hard state at 8.4\,GHz (Stirling et al. 2001).
This jet is also not as dramatic (in terms of rapid variability and
relativistic outflows) as in the cases of jets observed in
a number of other microquasars
e.g.  Cyg\,X-3 (Miouduszeski et al 2001), GRS\,1915+105 (Dhawan et al. 2000),
IE\,1740.7$-$2942 (Mirabel et al. 1992), and GRS\,1758-258 (Rodr\'{\i}guez et al. 1992).
Observations by Gallo et al (2005), however, suggest that the jet may be 
inflating a large (5\,pc) shell-like structure around the X-ray source,
implying that its contribution to the outflow of energy in this 
system is dominant.
Fig. 3 shows the low-frequency flux density from GMRT
plotted against the closest available 15-GHz flux density.
Different markers are used for the two GMRT frequencies.
There is a clear correlation between the high- and low-frequency data,
somewhat complicated by the upper limits on some of the GMRT flux densities,
but nevertheless convincing.

\subsubsection{The radio and X-ray association of Cyg\,X-1} 
The 15-GHz radio light curve and the RXTE/ASM X-ray count-rate and
hardness ratio HR2 for Cyg\,X-1
from 2003 Mar to 2005 Jan are shown in Fig.4
(daily averages for both of the X-ray parameters are used). The X-ray 
state of the system changed several times over this interval.
For the majority of the time, Cyg\,X-1 was in the low-hard state,
but with three distinct periods of the high-soft state lasting 1 -- 2 months each.
The suppression of the 15-GHz flux density during the high-soft states
is well-illustrated by these data.

There was one exceptional flare recorded in the 15-GHz data,
on 2004 Feb 20.27,
where the flux density reached a maximum of 140 mJy
in an event which lasted about 30 min.
Unfortunately there was no GMRT observation at this date.
\begin{figure*}[ht]
\begin{center}
\psfig{figure=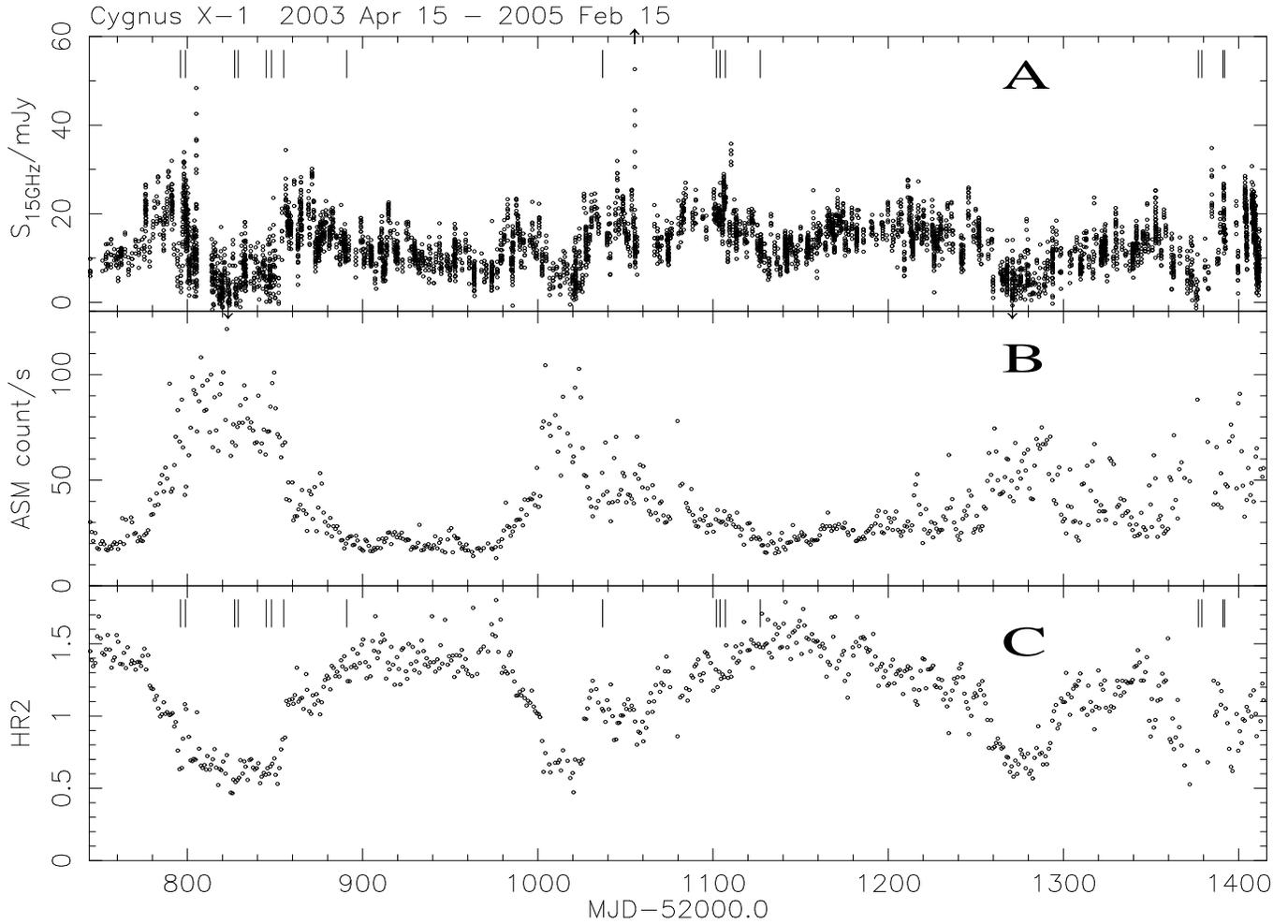,angle=0,width=18cm,height=13cm}
\end{center}
\caption{Cygnus\,X-1: Radio (15\,GHz) (top) and RXTE/ASM (middle) light-curves;
  ASM hardness ratio HR2=(5--12)keV/(3--5)keV (bottom)
  during the interval MJD 52796 = 2003 Jun 06 to MJD 53392 = 2005 Jan 22.
  The dates of GMRT observations are marked with vertical lines. }
\end{figure*}

\subsubsection{Spectral and temporal behaviour of Cyg\,X-1} 
Cyg\,X-1 is a persistent but variable flat-spectrum source over the range 2 to 220\,GHz
(Fender et al. 2000). It is clear from table 1 and Fig. 3 that Cyg\,X-1 was detected at low frequency in
only one of the 8 observations while it was in the high-soft X-ray state.
Observations in the intermediate and low-hard X-ray states
all resulted in  detections,
apart from two which both had relatively high noise levels.
Therefore we conclude that the high-soft state suppresses the
radio emission at low radio frequencies as it does at high frequencies.
There is no clear distinction between the
radio spectral index of the intermediate and low-hard states,
being typically $\alpha = 0.3$.
While all of these spectral measurements
are subject to the additional uncertainty from the lack of
precisely simultaneous observations, we also conclude that the radio
spectrum rises from the GMRT observing frequencies to
the regime where it is essentially flat (2 to 220\,GHz).
Part of the variation of radio emission from Cyg\,X-1 is associated
with the X-ray state of the system, and is therefore intrinsic.
A further part may be associated with
propagation through the interstellar medium --
in particular, refractive interstellar scintillation (RISS)
may introduce amplitude variations (Rickett, 1990).
We have used the NE2001 code made available by Cordes \& Lazio (1993)
to estimate the propagation parameters for Cyg\,X-1 and Cyg\,X-3.

The NE2001 code estimates that the transition frequency
(above which the scattering is weak, and therefore the intensity fluctuations small)
is around 6\,GHz. The angular broadening is estimated as
0.3\,mas at 0.61\,GHz and 0.06\,mas at 1.28\,GHz.
The scattering disc is therefore about $1 \times 10^{11}$m at 0.61\,GHz
and $2 \times 10^{10}$m at 1.28\,GHz. If we assume that typical transverse velocities
involved (the source or the medium) are near 100\,${\rm km\, s^{-1}}$, we derive
timescales of 5 and 1 days at the two frequencies.
These are clearly in the same area as those of the observed fluctuations,
so it is quite possible that RISS is a contributing factor.
We note, however, two further points.
The angular size of the (transient) radio jet mapped at 8.4\,GHz
by Stirling et al. (2001) is 10 mas: the low-frequency
emission may originate in a region of this size (or larger);
and the scintillation bandwidths estimated by the NE2001 code
are 30 and 100\,MHz at 0.61, 1.28\,GHz.
Both of these would reduce the observed  RISS contribution
to the fluctuations.

\subsubsection{Proposed geometry and emission mechanism}
From the above discussion, we conclude that the radio spectrum of Cyg\,X-1  consists of
two parts, the persistent high frequency emitting component and the variable
low frequency  component. Both  of these components imply the presence of a
continuous jet with discrete plasmoids or clouds of electrons, emitting synchrotron photons.
The observed behaviour of the source at radio frequencies can be explained by
adiabatically expanding plasmoids in conical jets (Hjellming et al. 1988),
 where the low frequency radio emitting component
is significantly altered by refractive interstellar scintillation giving rise to high
variability in the radio flux density.
A sketch of the proposed geometry is shown in Fig. 5. In our model
compact plasmoids  are ejected  continuously at the base of the jet and expand freely outward.
The low frequency emission from the  newly ejected plasmoids will be absorbed due to  synchrotron
self absorption  in an optically thick medium because of its highly compact nature and high magnetic field.
The plasmoid will be still visible  at high radio frequencies
(Miller-Jones et al. 2004). As the plasma expands under its internal pressure, 
it will become optically thin and emit over the whole 
radio-frequency band, although at a lower intensity.

The observed flat radio spectrum of  Cyg\,X-1 can thus be explained as due to
superposition of individual self-absorbed synchrotron components  emitted or plasmoids with differing
age profile and having longer life time. If a combination of size and  the magnetic field are such that
opacity effects are important only for the younger population, an inverted (or overall flat) spectrum is
seen. The average spectrum can also mimic a  flux level shift above a cut-off frequency,
characterized by the optical thickness of the medium and the choice of age distribution. 
\begin{figure}[h]
\begin{center}
\psfig{figure=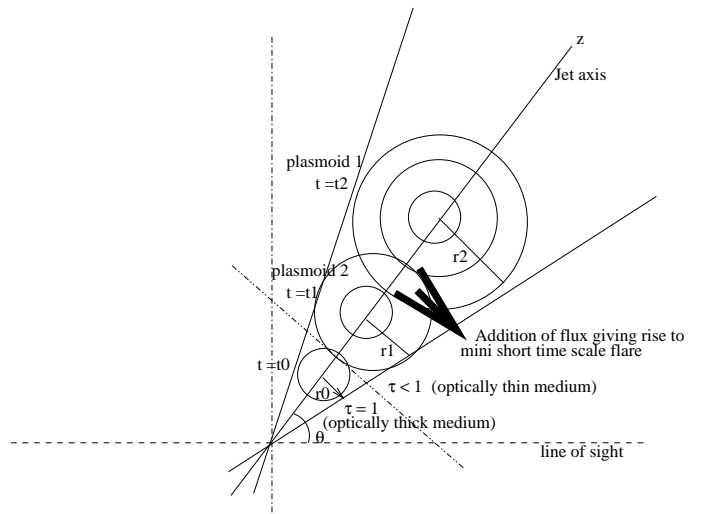,angle=0,width=9cm}
\end{center}
\caption{Model for Cyg\,X-1 jet}
\end{figure}
As discussed earlier, the 15-GHz and GMRT flux densities 
are correlated; while we do not believe that the 15-GHz 
data are significantly affected by scintillation, this 
may not be the case at the GMRT frequencies and 
at least some of the variance in the 0.61 and 1.28\,GHz data results form RISS. 
In conclusion, the radio emission in Cyg\,X-1 is mainly due to  a continuous
outflow of compact plasmoids or electron clouds into a jet and the spectrum consists of the composite
emission from each of these slow decaying ejecta and mini flaring events as observed in the source
 may result due to randomness of the age distribution of the ejected clouds.

\subsection{Cyg\,X-3}
Cyg\,X-3 is an X-ray binary system which does not seem to fit easily into
the established classes of these systems.
It is probably a low mass X-ray binary system. It has a 4.8-hr orbital period,
and is located in the Galactic plane at a distance of $\sim$10 kpc
(Predehl et al. 2002; Dickey 1983; Mason et al. 1979).
The optical counterpart of the X-ray source is a  WR star (van Kerkwijk et al. 1996),
which is not visible in the optical band because of heavy interstellar extinction but is clearly seen
above $\sim 0.8 {\rm \mu m}$.
The optical data also shows a 4.8-hr modulation corresponding  to the orbital
period as seen in X-rays  (Hanson et al. 1999).  The source exhibits two spectral states low-hard and
high-soft similar to Cyg\,X-1 and has been observed up to 500 keV during the flare mode. The quasi-periodic
oscillations with periods between 50 $--$ 1500  s are the key characteristics of the source
(van der Klis \& Janson 1985).  A detailed analysis of the X-ray spectrum suggests that the total no of
X-ray photons seems to be conserved at all times irrespective of the state and the observed spectrum is
consistent with a thermal source embedded in a hot plasma and enveloped in a cold hydrogen
 shell (Manchanda 2002).
The X-ray light curve of the source  in the 2-12 keV band  from the RXTE/ASM  data shows frequent
spectral changes between the high-soft to low-hard states  thereby suggesting
large changes in the accretion rate on to the compact object.

At radio wavelengths, Cyg\,X-3  is the most luminous X-ray binary in both its quiescent and
flaring states (Waltman et al. 1995).
Huge radio outbursts have been reported in Cyg\,X-3 during which the flux density can increase
up to levels of $\sim$20\,Jy; radio emission is suppressed (``quenched'')
to levels below 1\,mJy for some days before large radio flares (Waltman et al. 1994).
Jet-like structures with repeated relativistic ejection
have been observed at various radio frequencies (e.g. Schalinski et al. 1998).
On an arc-second scale,  two-sided jets have been seen from the source in the N-S orientation,
whereas a highly-relativistic ($\beta$ $\ge$ 0.81) one-sided jet with the same
orientation has been reported on milli-arcsec scales with the VLBA (Mart\'{\i} et al. 2001;
Mioduszewski et al. 2001).

\subsubsection{Temporal characteristics in the radio band and their association with  the X-ray emission}
In the Table 2  above we have summarized the radio flux densities of the source as  
measured during various observations with GMRT and from the Ryle Telescope data. Fig. 6 shows a 
flux--flux plot for Cyg\,X-3 for the GMRT and 15-GHz data,
and Fig. 7 shows the 15-GHz and RXTE ASM data for the whole period.
The timing of the GMRT observations is again marked with vertical lines in Fig. 7.
The last 4 GMRT observations, all at 0.61\,GHz,  were made during the high-soft state.
The mean flux density at that frequency was 33.2\,mJy, compared with 22.7\,mJy
for the 6 previous data points; the 15-GHz mean value was also higher, 150\,mJy
compared with 71\,mJy for the corresponding 6 data points. We note that the
radio/X-ray correlation is in the opposite sense to that for Cyg\,X-1.
McCollough et al (1999) report both anti-correlations (in the quiescent state)
and correlations (in the flaring state) between the hard X-ray flux
(20--100\,keV, as measured by BATSE) and the cm-wave radio flux density of Cyg\,X3.

We investigated the possibility that RISS might be important in the
case of Cyg\,X3. The propagation conditions are more severe than for the
case of Cyg\,X-1: the path length is much longer, and it has been known for some time
(e.g. Wilkinson et al, 1994) that the scatter-broadening for this source is extreme.
The NE2001 model is consistent: it suggests angular broadening of 20.59, 4.03\,arcsec
at 0.61, 1.28\,GHz. The corresponding timescales would be many years,
and the narrow scintillation bandwidth (2\,Hz) would suppress any observed scintillation.
We conclude that RISS is not relevant to this study of Cyg\,X-3.

\begin{figure}[h]
\begin{center}
\psfig{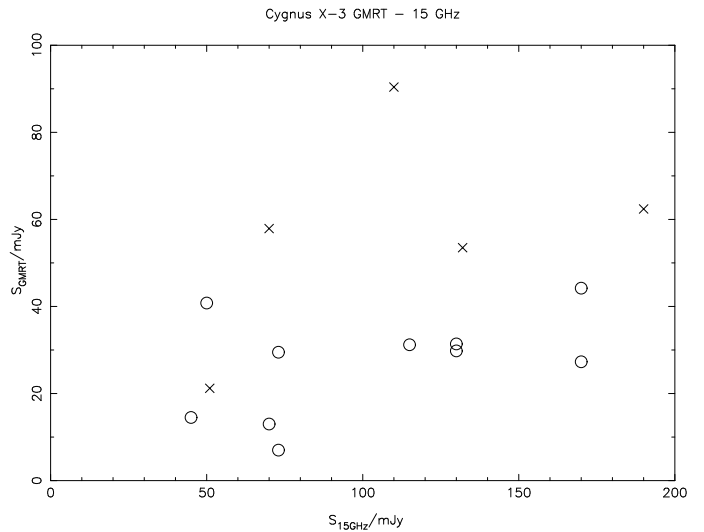}
\end{center}
\caption{Flux--flux plot for Cyg\,X-3 showing the low-frequency
GMRT flux densities ($\bigcirc$ 0.61\,GHz, $\times$ 1.28 GHz)
and the closest 15-GHz measurement}
\end{figure}

To look for correlation between the radio emission from the source
with  its  X-ray emission characteristics, we have plotted the  RXTE/ASM X-ray 
light curve for Cyg\,X-3  in Fig. 7 along with the radio data.
The timing of the GMRT observations is shown by the vertical lines.
It can be seen that no large flares were observed during this interval.
The radio emission is in the `quiescent' state, typically
50 to 200\,mJy at 15 GHz. For the last month or so of this time-range, the 
X-ray spectrum softens: the RXTE ASM ratio HR2 falls consistently below 2,
and the radio emission starts to become more erratic.
This behaviour faded away after another month or so,
and the source returned to the quiescent state.

\begin{figure*}[ht]
\begin{center}
\psfig{figure=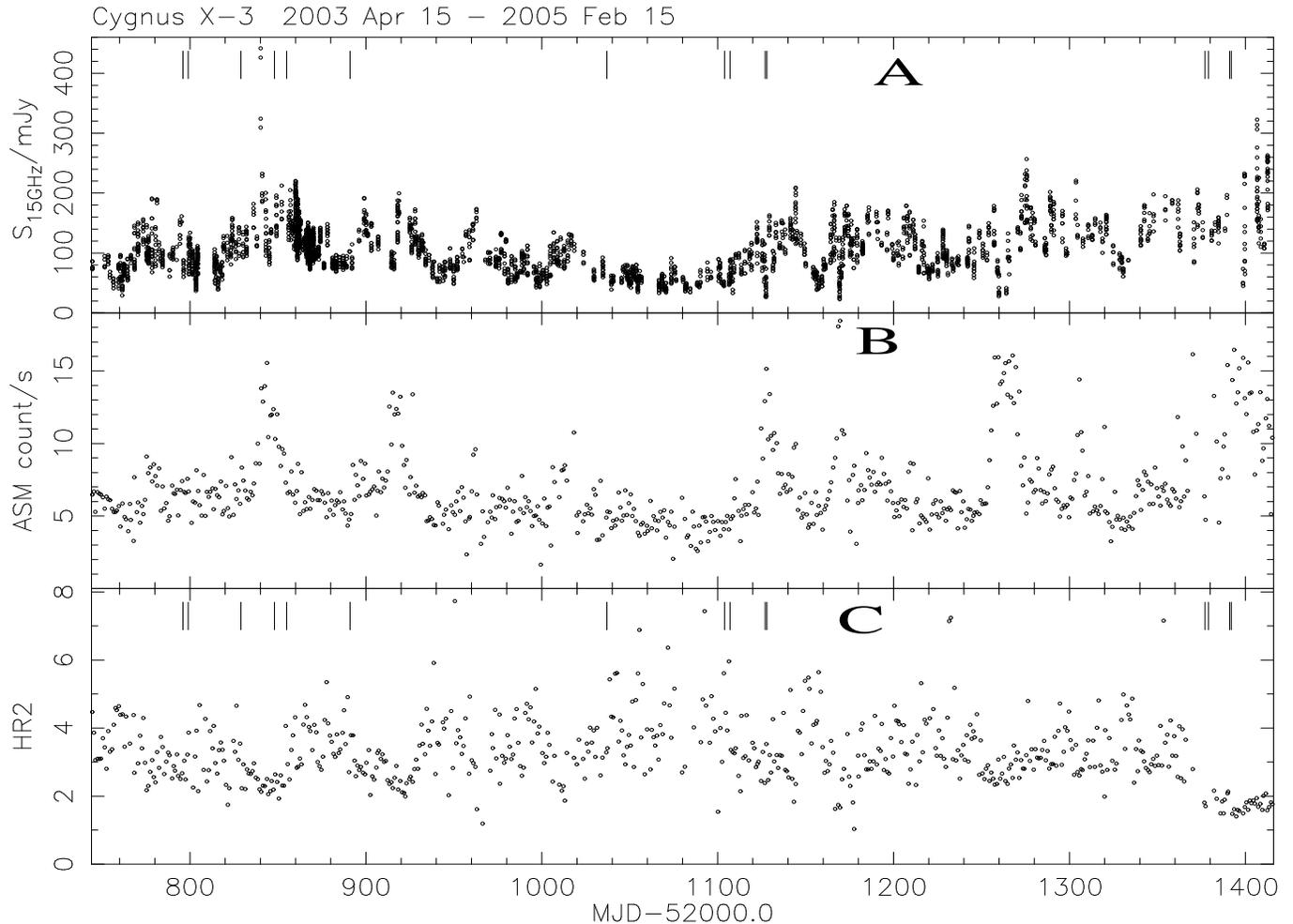,angle=0,width=18cm,height=13.0cm}
\end{center}
\caption{Cygnus\,X-3: Radio (15\,GHz) (top) and RXTE/ASM (middle) light-curves;
ASM hardness ratio (bottom), HR2=(5--12)keV/(3--5)keV during
the interval MJD 52796 = 2003 Jun 06 to MJD 53392 = 2005 Jan 22.
The dates of GMRT observations are marked with vertical lines.}
\label{fig.1.}
\end{figure*}

\subsubsection{Spectral behaviour and the emission geometry}
As seen from the data in Table 2 and Fig. 6, Cyg\,X-3 is a persistent radio source at all wavelengths. 
Cyg\,X-3  is more luminous at higher frequencies. The data in Table 2 clearly indicates a low frequency
turn-over in the source spectrum below 1\,GHz. As discussed earlier, such behaviour can arise due to
synchrotron self absorption of the compact radio emitting plasma in an optically thick medium. 
The observed variability of the flux density is consistent with the assumption of a discrete 
ejection/plasmoid in adiabatic expansion. The ejection rate and/or lifetimes of plasmoids in Cyg\,X-3 
are probably lower/shorter than in Cyg\,X-1 with little or no overlap of the spectra associated 
to the single discrete ejections. The spectrum and the fact that the radio emission is 
persistently at an high level may therefore imply an uniterrupted low-rate ejection.

\section{Summary}
In this paper we have presented the first ever low frequency maps of Cyg\,X-1 and Cyg\,X-3 
along with the temporal behaviour from the  monitoring observations of these sources over two 
years at various frequencies. It was noted that while Cyg\,X-3 is persistent source at low 
frequencies,  Cyg\,X-1 is transient in nature  with mean radio flux being 7\,mJy and the 
source is detected at low radio frequencies only when the high frequency radio flux crosses 
the mean threshold of 15\,mJy.  The spectral characteristics of both sources too have many 
similarities e.g. the spectral turn over at lower frequencies. Continuous blob of plasmoids 
are emitted within the jet medium in both the cases. However the life time of the individual 
blob in the case of Cyg\,X-1 is larger than in the case of Cyg\,X-3 and hence a flatness in
the radio spectra is seen in the case of Cyg\,X-1. The positive detection  of Cyg\,X-1 at 0.61 
to 1.2\,GHz on several occasions along with the spectral turnover at low frequencies conclusively 
support the synchrotron origin of the radio photons contrary to its ambiguous thermal or non-thermal 
nature concluded from earlier measurements (Fender et al. 2000). In the case of Cyg\,X-3, a close 
correlation is seen in the change of X-ray hardness ratio and  the radio flux densities during 
the flaring and quiescent stages thus providing a consistent picture of disc-jet connection.

Rigorous calculations were performed to look for the effect of refractive interstellar scintillation
at low frequency on the data of these two sources. The present data clearly indicate that
the source visibility above a threshold luminosity seen in Cyg\,X-1 is due to this effect.
However the variability at high frequencies is intrinsic to the source and least affected by
interstellar scintillations. Finally, we have discussed a common jet emission model for both the sources
in which synchrotron radiation-emitting plasmoids originate at the base of the conical jet and
the observed spectral features are due to the superposition of many such plasmoids with different
age profile. Thus detailed spectral measurements in the m-cm regime (low frequency radio) combined with
simultaneous X-ray data can provide a clear understanding of the emission mechanisms in these sources.
\begin{acknowledgement}
 We wish to thank the help received from  the GMRT staff during the observations.
MP wishes  to thank NCRA  for providing hospitality to carry on this work and  acknowledges the
financial support received from  Raman Research  Institute Trust, Bangalore during the early
part of this work.
She is also thankful to Prof. V. Kulkarni from Mumbai Univ. for constant encouragement.
Part of the work is supported by the Indian Space Research Organization grant under the REPOND program.
The RXTE ASM quick-look results provided by the ASM/RXTE team are gratefully acknowledged.
The Ryle Telescope is supported by PPARC.
\end{acknowledgement}

\end{document}